\documentclass[aps,11pt,notitlepage]{revtex4-1}
\usepackage{amsmath,amssymb,graphicx}
\usepackage[T1]{fontenc}
\usepackage[colorlinks=true, urlcolor=blue, citecolor=black]{hyperref}
\usepackage[font={small},flushleft,indent]{caption}

\begin{document}
\title{Spatial variation of the fine-structure constant and Hubble's law in anisotropic coordinate of Friedmann-Lemaître-Robertson-Walker space-time}
	
\author{Zhe Chang}
\author{Qing-Hua Zhu}
\email{zhuqh@ihep.ac.cn}

\affiliation{Institute of High Energy Physics, Chinese Academy of Sciences, Beijing 100049, China}
\affiliation{University of Chinese Academy of Sciences, Beijing 100049, China}

\begin{abstract}
Recent updated results of quasar spectra suggested a 3.9$\sigma$ significance of spatial variation of the fine-structure constant. Theoretically, it is important to examine whether the fine-structure constant, as a fundamental constant in quantum theory, is possible varying with space and time. In this paper, we explore the possibility that spatial variation of the fine-structure constant could be compatible with Einstein's general relativity. Namely, the spatially dependent fine-structure constant in the Universe could be originated in different values of the speed of light in separate local frames that are far away from us, since we have known that light rays must be bending in the present of gravity or non-inertial motions. In addition, to learn more about the anisotropic coordinate of FLRW space-time, we also study luminosity distance-redshift relation. It is found that there is a dipole structure in high redshift regime, while  in low redshift regime, there is not such dipole. 
\end{abstract}

{\maketitle}

\section{Introduction}

Recent updated results of quasar spectra suggested a 3.9$\sigma$ significance of spatial variation of the fine-structure constant \cite{wilczynska_four_2020}. In the last
decades, the varying fine-structure constant has been studied via observation on quasar spectra \cite{webb_indications_2011,webb_search_1999,wilczynska_four_2020,petitjean_constraining_2009,king_spatial_2012}, big bang nucleosynthesis \cite{mosquera_new_2013,avelino_early-universe_2001},
cosmic microwave background \cite{ ade_planck_2014,avelino_early-universe_2001,obryan_constraints_2014,de_martino_constraining_2016,balcerzak_modelling_2017,hart_new_2018,smith_probing_2019}, galaxy clusters \cite{de_martino_new_2016,de_martino_constraining_2016,colaco_galaxy_2019} and supernovae \cite{amendola_variation_2012,balcerzak_modelling_2017,li_testing_2019,amendola_variation_2012,leal_fine-structure_2014,murphy_subaru_2017,alves_current_2018}. Though it is accompanied by controversial \cite{obryan_constraints_2014,avelino_early-universe_2001,mosquera_new_2013,songaila_constraining_2014,whitmore_impact_2015,murphy_subaru_2017,martins_status_2017,colaco_galaxy_2019}, one might not doubt the importance of the fine-structure constant, as a fundamental constant, that is possible varying with space and time.

Confronted with variation of the fine-structure constant, Bekenstein considered a modified Maxwell's electromagnetism \cite{bekenstein_fine-structure_1982}, in which electron charge was assumed to be varying with space and time. It was generalized by Sandvik, Barrow and Magueijo \cite{barrow_variations_2002,sandvik_simple_2002,magueijo_is_2002}, the so-called Bekenstein-Sandvik-Barrow-Magueijo (BSBM) model, in which dynamical evolution of the fine-structure constant was involved in Einstein field equations by considering a scalar field coupled to electromagnetism. 
Recently, this framework was developed though considering evolution of perturbation \cite{barrow_gauge-invariant_2003}  and  modified evolution equation for the scalar field \cite{mota_local_2004,barrow_generalized_2012,barrow_general_2013,kraiselburd_magnetic_2018}. 
Besides, there are models focusing on physical aspect of the scalar field. Namely, the scalar field can be originated from a dilaton charge of atomic
systems \cite{damour_fast_2010,alexander_cyclic_2016}, quintessence type field of dark energy
\cite{amendola_variation_2012,martins_status_2017,boas_distinguishing_2020}, or dark matter \cite{silva_spatial_2014,davoudiasl_variation_2019}. In order to interpret variation of the fine-structure constant, there are also models suggesting a violence of the weak equivalence principle or Einstein equivalence principle \cite{bekenstein_fine-structure_1982,barrow_variations_2002,magueijo_is_2002,sandvik_simple_2002,damour_fast_2010,barrow_varying_2010,chang_fine_2012,hees_breaking_2014,martins_status_2017,davoudiasl_variation_2019,li_testing_2019,boas_distinguishing_2020}.

In this paper, we explore the possibility that spatial variation of the fine-structure constant could be compatible with Einstein's general relativity. It is motivated by the theory of varying speed of light \cite{barrow_varying-_1998,albrecht_time_1999,barrow_cosmologies_1999,magueijo_is_2002,barrow_varying_2010,salzano_probing_2016,alexander_cyclic_2016,salzano_statistical_2017}, and the fact that light rays are bending  in the present of gravitational fields or non-inertial motions. From Einstein equivalence principle, the physical laws of non-gravity theory should be the same as that in Minkowski space-time in the near region around geodesic observers. It means that the speed of light should be a constant in the local frame of the geodesic observers. Based on this, we suppose that there are two local frames with respect to geodesic observers $A$ and $B$ that are far away from each other. Both observers $A$ and $B$ could perceive a constant speed of light in their local frames. But it does not suggest that the values of their speed of light should be the same. Conversely, we have known that light rays must be bending in the present of gravity or non-inertial motions. Thus, the spatial variation of the fine-structure constraint might be originated in different values of the speed of light in separate local frames, as  the formula of the fine-structure constant $\alpha = \frac{e^2}{4\pi \epsilon_0  \hbar c}$ is always true in these local frames. 

In Minkowski space-time, inertial motion of reference frame would not change the metric of the space-time. This is suggested by Lorentz symmetry. However, in Friedmann-Lemaître-Robertson-Walker (FLRW) space-time, there is no longer such symmetry for geodesic reference frame.  It would lead to an anisotropic space-time for the frame adapted to geodesic observers \cite{chang_reference_2020}. In this paper, we study spatial variation of the fine-structure constant and luminosity-distance relation in the anisotropic coordinate of FLRW space-time. For the former, we will introduce a phenomenological definition for non-local measurement of speed of light  based on dataset of quasar spectra \cite{king_spatial_2012}. For the latter, it is the extension of the previous study \cite{chang_reference_2020} in high redshift regime.

This paper is organized as follows. In section~\ref{II}, we brief review the anisotropic coordinate of FLRW space-time. It is a global reference frame describing a congruence of observers that undergo geodesic motion along a specific direction. In section~\ref{III}, we study spatial variation of the fine-structure constant in the anisotropic FLRW space-time based on dataset of quasar spectra \cite{king_spatial_2012}. In section \ref{IV}, we present luminosity distance-redshift relation in the anisotropic FLRW space-time in general. For the sake of intuitive, we consider the luminosity distance-redshift relation in de Sitter and $\Lambda$CDM Universe as examples. Finally, conclusions and discussions are summarized in section~\ref{V}.

\section{Anisotropic coordinate of FLRW space-time} \label{II}

In the view of comoving observers in the Universe, the space-time can be described by FLRW metric, i.e.,
\begin{equation}
  {\rm{d}} s^2 = - {\rm{d}} t^2 + a^2 (t) (({\rm{d}} x^1)^2 + ({\rm{d}} x^2)^2 +   ({\rm{d}} x^3)^2) ~, \label{1}
\end{equation}
where scale factor $a (t)$ describes the expansion of the Universe. Different forms of $a (t)$ depend on cosmological models or epochs of the Universe. In this case, the Universe is isotropic and homogenous.

The picture of the Universe would be different, if we consider the Universe in the view of a congruence of observers that undergo geodesic motion along a specific direction  in FLRW space-time. Namely, for a coordinate system adapted to the geodesic observers, the metric is shown to be deviated from isotropy,
\begin{equation}
  {\rm{d}} s^2 = - {\rm{d}} T^2 + a^2 (T, Z) \left( {\rm{d}} X^2 + {\rm{d}} Y^2 +  \frac{1 + \left( \frac{\upsilon}{a} \right)^2}{1 + \upsilon^2} {\rm{d}} Z^2  \right) ~,\label{2}
\end{equation}
where the $\upsilon$ is an integral constant of 4-velocities of these geodesic observers.  This was also shown in previous study \cite{chang_reference_2020}.
The transformation between the coordinate systems $(T,X,Y,Z)$ and the isotropic FLRW space-time $(t,x^1,x^2,x^3)$ is given by
\begin{subequations}
\begin{eqnarray}
    T & = & \int_{t_0}^t \sqrt{1 + \left( \frac{\upsilon}{a (t')} \right)}
    {\rm{d}} t' - \upsilon x^3 ~,\\
    X & = & x^1 ~,\\
    Y & = & x^2 ~,\\
    Z & = & \sqrt{1 + \upsilon^2} \left( x^3 - \upsilon \int_{t_0}^t
    \frac{{\rm{d}} t'}{a^2 (t') \sqrt{1 + \left( \frac{\upsilon}{a (t')}
    \right)^2}} \right) ~, 
\end{eqnarray}\label{3}
\end{subequations} 
where we have adopted the conventions that $t=t_0$ and $T=0$ are the time of the present Universe.  In the following, we will call the coordinate system $(T, X, Y, Z)$ as anisotropic (coordinate of) FLRW space-time. 

From Eqs.~(\ref{3}), one might find  that the coordinate transformation returns Lorentz transformation as $a \rightarrow 1$. Based on the transformation, the 4-velocities of comoving observers $u = \partial_T$ in the anisotropic FLRW space-time can be transformed into $u^{\mu} = \sqrt{1 + \left( \frac{\upsilon}{a} \right)^2} \partial_t + \frac{\upsilon}{a^2} \partial_{x^3}$ that exactly describes the geodesic observers in the isotropic FLRW space-time.

By making use of Eq.~(\ref{3}), we can obtain the scale factor $a$ via solving
\begin{eqnarray}
  T + \frac{\upsilon}{\sqrt{1 + \upsilon^2}} Z & = & \int_1^a \frac{{\rm{d}}
  a}{H \sqrt{a^2 + \upsilon^2}} ~, \label{4}
\end{eqnarray}
where $H~(\equiv {a^{-1}} {{\rm{d}} a}/{{\rm{d}} t})$ is Hubble parameter. In the anisotropic coordinate of FLRW space-time, the scale factor $a$ has been expressed in terms of coordinates $T$ and $Z$. Thus the Universe described by the metric in Eq.~(\ref{2}) is beyond the Bianchi-type Universe \cite{Ellis:1998ct}. 

\section{Spatial variation of the fine-structure constant in anisotropic FLRW space-time} \label{III}

The fine-structure constant $\alpha$ is known as a
fundamental constant in quantum theory. In Minkowski space-time, it can be expressed in terms of electron charge $e$, vacuum permittivity $\epsilon_0$, Planck constant $\hbar$ and the speed of light $c$,
\begin{equation}
  \alpha = \dfrac{e^2}{4\pi \epsilon_0 \hbar c} ~.
\end{equation}
However, as suggested by Webb \emph{et~al.}~\cite{webb_indications_2011}, the fine-structure constant was
found to be spatially dependent  based on observation of quasar spectra in the
Universe. Recently, the results of spatial variation of the fine-structure constant was updated
and confirmed again with significance 3.9$\sigma$ \cite{wilczynska_four_2020}.

As suggested by Einstein equivalence principle, local measurements of speed of light should always be a constant. However, it does not mean that the constant speed of light in separate local frames have the same values in the present of gravity. Therefore, we argue that the variation of the fine-structure constant could be originated in the speed of light affected by gravity or non-inertial motion within Einstein's relativity. In this sense, we could formulate the variation of the fine-structure constant in the form of
\begin{equation}
  \frac{\Delta \alpha}{\alpha} \equiv \frac{\alpha (z) - \alpha}{\alpha} =
  \frac{1}{c_z} - 1 ~, \label{7}
\end{equation}
where we utilize convention $c = 1$ for the speed of light on the earth, and the $c_z$ is the speed of light in the place far from us, and thus can not be measured directly. In this sense, resolving the variation of the fine-structure constant in the Universe turns to be how to define non-local measurement of speed of light $c_z$.

In the following, we will present a phenomenological formula for the speed of light $c_z$. As baseline, we start from Milne Universe in which the speed of light should be a universal constant.

\subsection{Fine-structure constant in Milne Universe}

In the early stage of cosmology development, Milne proposed a model of the expanding
Universe within special relativity \cite{mukhanov_physical_2005}. The metric of the space-time is given by
\begin{equation}
  {\rm{d}} s^2 = - {\rm{d}} t^2 + (C  t)^2 \left( \frac{{\rm{d}} r^2}{1 + (C
   r)^2} + r^2 ({\rm{d}} \theta^2 + \sin^2 \theta {\rm{d}} \phi^2) \right)
  ~,\label{8}
\end{equation}
where $C$ is a constant with dimension of 1/length and $t > 0$. In this model, the metric in Eq.~(\ref{8}) can
be transformed into Minkowski space-time via simply a coordinate
transformation. Although, it has been falsifiable by observation on expansion rate of the Universe, such as supernovae \cite{riess_observational_1998} for example, it could still provide insights for understanding the Universe.

Since comoving observers in Milne Universe undergo inertial motion (i.e., uniform motion), the speed of light should be a universal constant in the view of all the comoving observers. In this point of view, we suggest that the speed of light $c_z$ could be formulated as
\begin{equation}
  c_z \equiv \lim_{\Delta t \rightarrow 0} \frac{1}{\Delta t} \left(
  \frac{\Delta d_{\rm c}}{1 + z_{\rm knt}} \right) = \frac{1}{1 + z_{\rm knt}} \frac{{\rm{d}}  d_{\rm c}}{{\rm{d}} t}
  ~, \label{9}
\end{equation}
where the comoving distance $ d_{\rm c}$ is defined as
\begin{eqnarray}
   d_{\rm c} & \equiv & \int \sqrt{\gamma_{i  j} {\rm{d}} x^i {\rm{d}} x^j}
  \nonumber\\
  & = & \int \left( \frac{{\rm{d}} r^2}{1 + (C  r)^2} + r^2 ({\rm{d}}
  \theta^2 + \sin^2 \theta {\rm{d}} \phi^2) \right)^{\frac{1}{2}} ~.
\end{eqnarray}
Here, the $\gamma_{i j}$ is conformal metric of $g_{i j}$.  In order to obtain a proper speed of light $c_z$,  we argue that the comoving distance $d_{\rm c}$ should be re-adjusted by a kinematic redshift $1 + z_{\rm knt}$ in Eq.~(\ref{9}). 

In the following, we will check it in detail that there is a universal constant speed of light  in Milne Universe. We consider the 4-velocities of radial light rays in the form of
\begin{subequations}
\begin{eqnarray}
		\frac{{\rm{d}} t}{{\rm{d}} \lambda} & = & \frac{1}{C  t} ~,\\
		\frac{{\rm{d}} r}{{\rm{d}} \lambda} & = & \frac{\sqrt{1 + (C 
				r)^2}}{(C  t)^2} ~,\\
		\frac{{\rm{d}} \phi}{{\rm{d}} \lambda} & = & 0 ~,\\
		\frac{{\rm{d}} \theta}{{\rm{d}} \lambda} & = & 0 ~.
\end{eqnarray} \label{11}
\end{subequations}
Based on Eqs.~(\ref{11}), we obtain observed redshift $z$ of comoving objects
\begin{equation}
  1 + z = \frac{- u_{\nu} \frac{{\rm{d}} x^{\nu}}{{\rm{d}} \lambda}
  \Big|_{\rm{emit}}}{ - u_{\mu} \frac{{\rm{d}} x^{\mu}}{{\rm{d}}
  \lambda} \Big|_{\rm{obs}}} = \frac{1}{C  t} ~. \label{12}
\end{equation}
Since expansion of the Milne Universe is a pure inertial effect, it leads to
\begin{equation}
	z=z_{\rm knt}. \label{12-1}
\end{equation}
By making use of Eqs.~(\ref{9}), (\ref{12}) and (\ref{12-1}), we obtain the speed of light $c_z$ explicitly
\begin{eqnarray}
  c_z & = & \frac{1}{1 + z} \sqrt{\frac{1}{1 + (C  r)^2} \left(
  \frac{{\rm{d}} r}{{\rm{d}} t} \right)^2 + r^2 \left( \frac{{\rm{d}} \theta}{{\rm{d}}
  t} \right)^2 + r^2 \sin^2 \theta \left( \frac{{\rm{d}} \phi}{{\rm{d}} t}
  \right)^2} \label{13}\nonumber\\
  & = & 1 ~.
\end{eqnarray}
It shows that the $c_z$ is a universal constant, and can describe the speed of
light in the view of comoving observers in Milne Universe. Thus, the formula of $c_z$ in Eq.~(\ref{9}) is acceptable according to the baseline testing in Eq.~(\ref{13}).

Therefore, based on Eq.~(\ref{13}), there is no variation of the
fine-structure constant, namely,
\begin{equation}
  \frac{\Delta \alpha}{\alpha} = \frac{1}{c_z} - 1 = 0 ~.
\end{equation}


\subsection{Fine-structure constant in isotropic FLRW space-time}

In the present of gravity, the kinematic part of the redshift $z_{\rm{knt}}$ turns to be
\begin{eqnarray}
  1 + z_{\rm{knt}} & = & \frac{1 + z}{1 + z_g} ~, \label{16}
\end{eqnarray}
where the $z$ and the $z_g$ are cosmological redshift and gravitational redshift, respectively. In this sense, we assume that the observed redshift $z$ is originated from two different contributions, the kinematic effect and the gravitational effect. And both the $z_{\rm{knt}}$ and the $z_g$ can not be observed, directly.  

By making use of Eq.~(\ref{16}), we will calculate the speed of light $c_z$ in isotropic FLRW space-time in the following. 
From null geodesic equations, we obtain the 4-velocities of light rays, i.e.,
 \begin{subequations}
 	\begin{eqnarray}
 		\frac{{\rm{d}} t}{{\rm{d}} \lambda} & = & \frac{\sqrt{\delta_{i  j} l^i
 				l^j}}{a} ~,  \\
 		\frac{{\rm{d}} x^i}{{\rm{d}} \lambda} & = & \frac{l^i}{a^2} ~, \label{18}
 	\end{eqnarray}\label{17}
 \end{subequations}
where $l^i$ are integral constants from solving the geodesic equations. Thus, the observed redshift $z$ of co-moving objects is given by
\begin{equation}
  1 + z = \frac{ - u_{\nu} \frac{{\rm{d}} x^{\nu}}{{\rm{d}} \lambda}
 \Big|_{\rm{emit}}}{ - u_{\mu} \frac{{\rm{d}} x^{\mu}}{{\rm{d}}
  \lambda} \Big|_{\rm{obs}}} = \frac{1}{a} ~, \label{19}
\end{equation}
where $u^{\mu}~( = (1, 0, 0, 0))$ represents the 4-velocities of comoving objects.  Based on Eqs.~(\ref{16})--(\ref{19}), we obtain the speed of light $c_z$ in the form of
\begin{equation}
  c_z = 1 + z_g ~. \label{19-1}
\end{equation}
In the FLRW space-time, the speed of light $c_z$ has different value from the $c = 1$ at local. Here, the gravitational redshift $z_g$ might not be obtained, theoretically, because there seems not a theorem within general relativity that could decompose the gravitation and inertial effect, formally. 

Based on Eqs.~(\ref{7}) and (\ref{19-1}), the variation of the fine-structure constant is shown to be
\begin{equation}
  \frac{\Delta \alpha}{\alpha} = \frac{1}{1 + z_g} - 1 ~.
\end{equation}
 From the observation on quasar spectra \cite{webb_search_1999}, the variation of the fine-structure constant was shown to be $\frac{\Delta\alpha}{\alpha} \sim 10^{- 5}$. Based on this, we can infer $z_g \sim 10^{-5}$. It suggests that expansion of the Universe might be dominated by the kinematical redshift $z_{\rm knt}$, since $z_{\rm knt}\gg z_g$. 

\subsection{Fine-structure constant in anisotropic FLRW space-time}

From the observation on CMB \cite{ade_planck_2014} and quasar spectra \cite{webb_indications_2011,wilczynska_four_2020}, the Universe is shown to be not precisely isotropic. Although the observed anisotropy is very small, it is still interesting to consider what is the physical origins of the anisotropy of the Universe. In this part, we will study the anisotropy of the Universe from the observation on quasar spectra \cite{webb_indications_2011,wilczynska_four_2020}. Namely, the spatial variation of the fine-structure constant is originated in different values of the speed of light $c_z$ in the local frames of quasars. 

We also start our calculation from the speed of light $c_z$ based on Eqs.~(\ref{9}) and (\ref{16}). From the metric given in Eq.~(\ref{2}), 4-velocities of light rays can be obtained to be
\begin{subequations}
\begin{eqnarray}
	\frac{{\rm{d}} T}{{\rm{d}} \lambda} & = & \frac{\sqrt{(l_1)^2 + (l_2)^2 +
			(l_3)^2} \sqrt{a^2 + \upsilon^2} - \upsilon l_3}{a^2} ~,\\
	\frac{{\rm{d}} X}{{\rm{d}} \lambda} & = & \frac{l_1}{a^2} ~,\\
	\frac{{\rm{d}} Y}{{\rm{d}} \lambda} & = & \frac{l_2}{a^2} ~,\\
	\frac{{\rm{d}} Z}{{\rm{d}} \lambda} & = & \frac{\sqrt{1 + \upsilon^2}}{a^2}
	\left( l_3 - \frac{\upsilon \sqrt{(l_1)^2 + (l_2)^2 + (l_3)^2}}{\sqrt{a^2
			+ \upsilon^2}} \right) ~,
\end{eqnarray}
\end{subequations}
where $l_1$, $l_2$ and $l_3$ are integral constants from solving null geodesic equations. Thus, we can obtain observed redshift of comoving objects (sources) as follows,
\begin{eqnarray}
  1 + z & = & \frac{ - g_{\mu \nu} u^{\mu} \frac{{\rm{d}} X^{\mu}}{{\rm{d}}
  \lambda} \Big|_{\rm{src}}}{ - g_{\sigma \rho} u^{\sigma}
  \frac{{\rm{d}} X^{\rho}}{{\rm{d}} \lambda} \Big|_{\rm{obs}}} \nonumber\\
  & = & \frac{1}{a_{\rm{src}}^2} \frac{\sqrt{a_{\rm{src}}^2 + \upsilon^2}
  - \upsilon \gamma}{\sqrt{1 + \upsilon^2} - \upsilon \gamma} ~, \label{22} 
\end{eqnarray}
where $a_{\rm{src}} \equiv a (T_{\rm{src}}, Z_{\rm{src}})$, $X_{\rm{src}}^{\mu}$ is the event of a light ray emitted from the source, and we have redefined the integral constant $\gamma \equiv l_3 ((l_1)^2 + (l_2)^2 + (l_3)^2)^{-\frac{1}{2}}$. In order to locate the light rays observed in observers' celestial sphere, we can define an angle $\Theta$ as follows \cite{Soffel:2019aoq},
\begin{eqnarray}
  \cos \Theta & = & \left( \frac{k \cdot w}{(u \cdot k) (u \cdot w)} + 1
  \right)_{a = 1} \nonumber\\
  & = & - \frac{\gamma \sqrt{1 + \upsilon^2} - \upsilon}{\sqrt{1 +
  \upsilon^2} - \upsilon \gamma} ~,  \label{23}
\end{eqnarray}
where $k$ is a reference null vector with $\gamma = - 1$, and $w$ is the null vector of a light ray from the source. In the case of $\Theta=0^\circ$, it indicates that the source is located at positive $Z$-axis with respect to observers. 
By making use of  Eqs.~(\ref{22}) and (\ref{23}), and substituting $t\rightarrow T$ in Eq.~(\ref{9}), we obtain the speed of light $c_z$ in the form of
\begin{eqnarray}
   c_z & = & \frac{1 + z_g}{1 + z} \sqrt{\gamma_{i  j} \frac{{\rm{d}}
  X^i}{{\rm{d}} T} \frac{{\rm{d}} X^j}{{\rm{d}} T}} \nonumber\\
  & = & \frac{1 + z_g}{1 + z} \Bigg( \frac{1}{4 (1
  	+ z)^2} \Big(  \sqrt{\left( 2 \sqrt{1 + \upsilon^2} (1 + z) +
  \upsilon \cos \Theta - \sqrt{1 + \upsilon^2} \right)^2 - 4 (1 + z) z}     \nonumber\\
 & &    - \upsilon \cos \Theta + \sqrt{1 + \upsilon^2}  \Big)^2 - \upsilon^2 \Bigg)^{- \frac{1}{2}} ~.
\end{eqnarray}

Therefore, the variation of the fine-structure constant is formulated as
\begin{eqnarray}
  \frac{\Delta \alpha}{\alpha} & = &  \frac{1 + z}{1 + z_g} \Bigg( \frac{1}{4 (1
  	+ z)^2} \Big(  \sqrt{\left( 2 \sqrt{1 + \upsilon^2} (1 + z) +
  	\upsilon \cos \Theta - \sqrt{1 + \upsilon^2} \right)^2 - 4 (1 + z) z}     \nonumber\\
  & &    - \upsilon \cos \Theta + \sqrt{1 + \upsilon^2}  \Big)^2 - \upsilon^2 \Bigg)^{ \frac{1}{2}} - 1 ~. \label{25}
\end{eqnarray}
It shows that the value of $\Delta \alpha / \alpha$ depends on spatial location $\Theta$ and the redshift  $z$ of the sources. In the expansion at  $\upsilon \rightarrow 0$, one can find a spatial dipole structure of the $\Delta \alpha/\alpha$, i.e.,
  \begin{eqnarray}
 \frac{\Delta \alpha}{\alpha}  =  \frac{1}{1 + z_g} - 1 + \left( \frac{z}{1 + z_g}
  \right) \upsilon \cos \Theta +\mathcal{O} (\upsilon^2) ~. \label{27-1}
\end{eqnarray}

\begin{figure}
	{\includegraphics[width=1\linewidth]{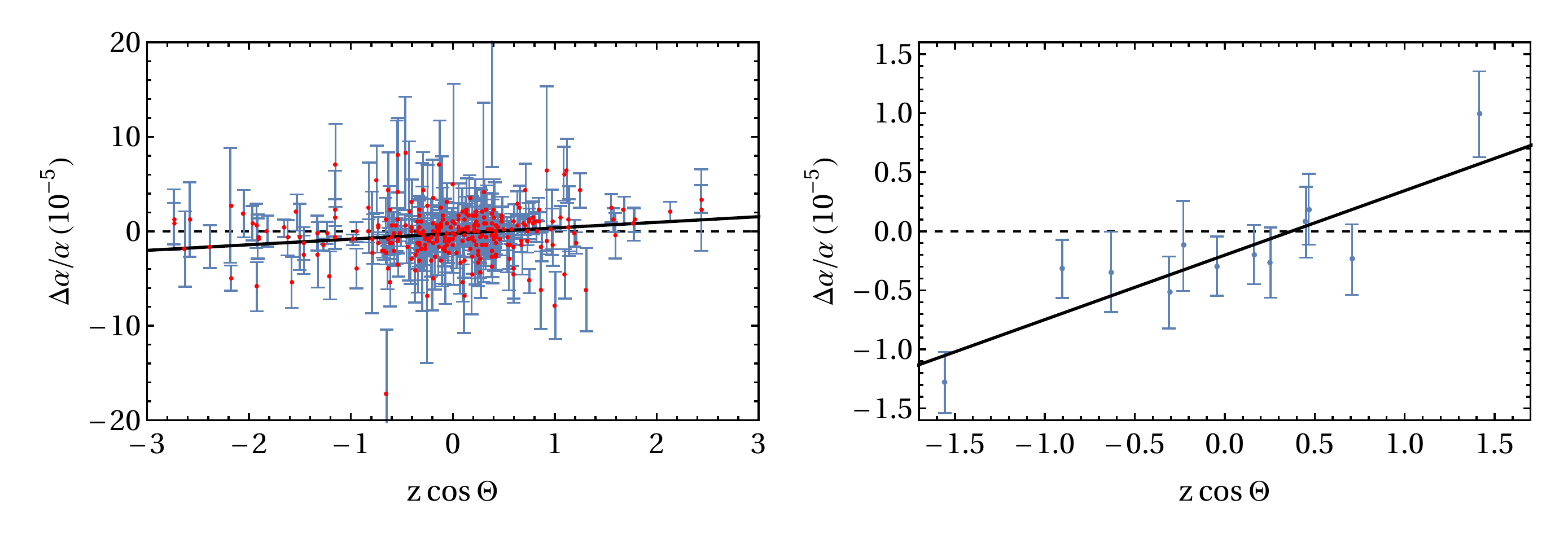}
	\caption{The value of $\frac{\Delta \alpha}{\alpha}$ plotted against $z\cos \Theta$. The $\Theta$ is the angle from the location of dipole given in Ref.~\cite{king_spatial_2012}. The solid lines are the best-fitting curves based on  Eq.~(\ref{27-1}).  Left panel: The data of quasar spectra are unbinned.  Right panel: The data of quasar spectra are binned.}\label{F1}}
\end{figure}

In Figure~\ref{F1}, we present plots of the values of $\frac{\Delta \alpha}{\alpha}$ against  $z \cos \Theta$ with available dataset of quasar spectra \cite{king_spatial_2012}. Here, we didn't remove any outliers as it did in Ref. \cite{king_spatial_2012}.
Based on these data,  best-fit curves of Eq.~(\ref{27-1}) are obtained by using weighted least squares method.  The parameters $\upsilon$ and $z_g$ in our fits are listed in Table~{\ref{T1}}. It shows that the value of $\upsilon$ is much smaller than the value of $\upsilon$ inferred from the redshift survey in previous study \cite{chang_reference_2020}. Since $z_g>0$ in our fits, it suggests existence of gravitational repulsion in the Universe. 
This is consistent with our present knowledge about the accelerating expansion of the Universe \cite{riess_observational_1998}
\begin{table}
	\caption{Fitting parameters of best-fit curves based on Eq.~(\ref{27-1}). }\label{T1}
 \begin{tabular}{ccccc}
 	\hline\hline
 	Fitting Parameters ($10^{-6}$) & \hspace{1cm} &unbinned  &\hspace{1cm} & binned\\
 	\hline
 	$z_g$			& &	$2.4 \pm 0.8$		&	& 		$2.0 \pm 0.8$	\\
 	$\upsilon/c$	& &		$5.9 \pm 1.2$	&	&	$5.5 \pm 1.1$		\\
 	\hline
 \end{tabular}
\end{table}

The results shown above are independent of cosmological models, but depend on the way that we  formate the $c_z$, i.e., Eq.~(\ref{9}). In principle, the formulation of $c_z$ indicates how we infer and calculate the  speed of light $c_z$ and should be based on the first principle. At the first step, here we  present a phenomenological formula of $c_z$. A better definition of measurement of the $c_z$ is expected to be found in future studies.

\section{Luminosity distance-redshift relation in anisotropic coordinates of FLRW space-time} \label{IV}

The luminosity distance $d_L$ from an object of luminosity $L$ and observed
flux $F$ is defined as \cite{dodelson_modern_2003,mukhanov_physical_2005}
\begin{equation}
  d_L \equiv \sqrt{\frac{L}{4 \pi F}} ~. \label{27}
\end{equation}
This definition in Eq.~(\ref{27}) 
is independent on cosmological model and is given phenomenologically. 

Since expansion of the Universe, the observed luminosity $L_{\rm{obs}}$ should be
smaller than the luminosity at the source $L$ by factor $(1 + z)^2$, namely
$L_{\rm{obs}} = (1 + z)^{- 2} L$. Therefore, based on definition of observed flux
$F$ through a closed isochronous surface $\mathcal{S}$, we have
\begin{equation}
  F\mathcal{S}= L_{\rm{obs}} = \frac{L}{(1 + z)^2} ~. \label{28}
\end{equation}
Using  Eqs.~(\ref{27}) and (\ref{28}), we obtain luminosity distance in the form of
\begin{equation}
  d_L = (1 + z) \sqrt{\frac{\mathcal{S}}{4 \pi}} ~.  \label{29}
\end{equation}
Finally, to obtain luminosity distance-redshift relation, the area of isochronous surface $\mathcal{S}$ should be expressed in terms of redshift $z$. 

In the following, by making use of luminosity distance in Eq.~(\ref{29}),  we will brief review luminosity
distance-redshift relation in isotropic FLRW space-time, and then extend this approach in the anisotropic coordinate of FLRW space-time.

\subsection{Brief review of Luminosity distance-redshift relation in isotropic FLRW space-time}

From 4-velocities of light rays in Eqs.~(\ref{17}), we obtain  comoving coordinates $x^\mu$ of the sources,
\begin{equation}
  x^k = \frac{l^k}{\sqrt{\delta_{i  j} l^i l^j}} \int_t^{t_0} \label{30}
  \frac{{\rm{d}} t}{a} ~,
\end{equation}
where $t$ is the time when light rays are emitted from the sources, $t_0$ is the  time of the present Universe. Based on
Eq.~(\ref{30}), we obtain equation of isochronous surface,
\begin{equation}
  \sqrt{(x^1)^2 + (x^2)^2 + (x^3)^2} = \int_t^{t_0} \frac{{\rm{d}} t}{a} =
  \int_0^z \frac{{\rm{d}} z}{H} ~. \label{31}
\end{equation}
At right hand side of above equations, the $\sqrt{(x^1)^2 + (x^2)^2 + (x^3)^2}$ is known as co-moving distance between the observers on the earth and source.

Using the surface equation in Eq.~(\ref{31}), we obtain the area of the isochronous
surface $\mathcal{S}$,
\begin{equation}
  \mathcal{S}= 4 \pi \left( \int_0^z \frac{{\rm{d}} z}{H} \right)^2 ~. \label{32}
\end{equation}
Therefore, based on Eqs.~(\ref{29}) and (\ref{32}), the luminosity distance-redshift
relation takes the form of
\begin{equation}
  d_L = (1 + z) \int_0^z \frac{{\rm{d}} z}{H} ~.
\end{equation}
This is the well-known luminosity distance-redshift relation in FLRW space-time \cite{dodelson_modern_2003,mukhanov_physical_2005}.

In the following, we will use this approach to calculate the luminosity distance-redshift relation in anisotropic FLRW space-time.

\subsection{Luminosity distance-redshift relation in anisotropic FLRW space-time \label{B}}

Since there is no difference in  $X$-axis and $Y$-axis, we let $l_2 = 0$ in Eqs.~(\ref{22}) for simplicity. Then, we can rewrite Eqs.~(\ref{22}) in the form of
\begin{subequations}
	\begin{eqnarray}
		\frac{{\rm{d}} T}{{\rm{d}} X} & = & \frac{\sqrt{a^2 + \upsilon^2} - \upsilon
			\gamma}{\alpha} ~, \label{34}\\
		\frac{{\rm{d}} Z}{{\rm{d}} X} & = & \frac{\sqrt{1 + \upsilon^2}}{\alpha} \left(
		\gamma - \frac{\upsilon}{\sqrt{a^2 + \upsilon^2}} \right) ~, \label{35}
	\end{eqnarray} \label{35-1}
\end{subequations}
where we have redefined the integral constant $\alpha \equiv l_1 / \sqrt{(l_1)^2 + (l_2)^2 + (l_3)^2}$. Based on
Eq.~(\ref{4}) and above equations, coordinate $X$ can be expressed in terms of
scalar factor $a$ along a worldline of light ray, i.e.,
\begin{equation}
  X - X_{\rm{src}} = \alpha \int^{a (T, Z)}_{a_{\rm{src}}} \frac{{\rm{d}}
  a}{H  a^2} ~, \label{36}
\end{equation}
where $a_{\rm{src}} \equiv a (T_{\rm{src}}, Z_{\rm{src}})$, and the
$(T_{\rm{src}}, Z_{\rm{src}})$ is the event at which light rays are
emitted from the source. By rewriting the Eq.~(\ref{36}) in form of $\frac{{\rm{d}}
X}{{\rm{d}} a} = \frac{\alpha}{H  a^2}$ and substituting it into
Eqs.~(\ref{35-1}), we obtain the worldlines of light rays, 
\begin{subequations}
\begin{eqnarray}
	Z - Z_{\rm{src}} & = & \frac{\gamma \hspace*{\fill}}{\alpha} \sqrt{1 +
		\upsilon^2} (X - X_{\rm{src}}) - \int^{a (T, Z)}_{a_{\rm{src}}} {\rm{d}} a
	\left\{ \frac{\upsilon}{a^2 H} \sqrt{\frac{1 + \upsilon^2}{a^2 +
			\upsilon^2}} \right\} ~,  \label{37}\\
	T - T_{\rm{src}} & = & - \frac{\upsilon \gamma}{\alpha} (X -
	X_{\rm{src}}) + \int^{a (T, Z)}_{a_{\rm{src}}} {\rm{d}} a \left\{
	\frac{\sqrt{a^2 + \upsilon^2}}{a^2 H} \right\} ~. \label{38}
\end{eqnarray} \label{36-1}
\end{subequations}
Based on Eqs.~(\ref{36-1}), we thus obtain the equation of isochronous surface of light rays from
the source at event $(T_{\rm{src}}, X_{\rm{src}}, Y_{\rm{src}}, Z_{\rm{src}})$,
\begin{equation}
  (X - X_{\rm{src}})^2 + (Y - Y_{\rm{src}})^2 + \left( \frac{Z -
  Z_{\rm{src}}}{\sqrt{1 + \upsilon^2}} + \upsilon
  \int^{a_0}_{a_{\rm{src}}} \frac{{\rm{d}} a}{a^2 H \sqrt{a^2 + \upsilon^2}}
  \right)^2 = \left( \int_{a_{\rm{src}}}^{a_0} \frac{{\rm{d}} a}{a^2 H}
  \right)^2 ~, \label{39}
\end{equation}
where $a_0 \equiv a (0, Z)$. Here, we have recovered the direction of $Y$ by simply adding the terms $(Y - Y_{\rm{src}})^2$ at the left hand side of above equation. In the case of $\upsilon \rightarrow 0$, one can check that it returns to Eq.~(\ref{31}).

In the anisotropic coordinate of FLRW space-time, the area of the isochronous
surface $\mathcal{S}$ is shown to be more complex. It can be obtained by
making use of Eq.~(\ref{39}), i.e.,
\begin{eqnarray}
  \mathcal{S} & = & 2 \pi \int^{Z_2}_{Z_1} {\rm{d}} Z \left\{ |X - X_{\rm{src}}| \sqrt{1
  + \frac{1 + \left( \frac{\upsilon}{a} \right)^2}{1 + \upsilon^2} \left(
  \frac{{\rm{d}} X}{{\rm{d}} Z} \right)^2} \right\} \nonumber\\
  & = & 2 \pi \int^{Z_2}_{Z_1} {\rm{d}} Z \left\{ \left( \left( \left( \frac{1
  + \left( \frac{\upsilon}{a_0} \right)^2}{1 + \upsilon^2} \right)^2 \left(
  \frac{\upsilon}{a_0} \right)^2 + 1 \right) A^2 + \left( \left( \frac{1 +
  \left( \frac{\upsilon}{a_0} \right)^2}{1 + \upsilon^2} \right)^2 \left( 1 +
  \left( \frac{\upsilon}{a_0} \right)^2 \right) - 1 \right) B^2 
\right. \right.   \nonumber\\
  & & \left. \left.- 2 \left( \frac{1 + \left( \frac{\upsilon}{a_0} \right)^2}{1 + \upsilon^2}
 \right)^2 \frac{\upsilon}{a_0} \sqrt{1 + \left( \frac{\upsilon}{a_0}
 \right)^2} A  B \right)^{\frac{1}{2}} \right\} ~, \label{40}
\end{eqnarray}
where the $A$ and $B$ are defined as
\begin{subequations}
	\begin{eqnarray}
		A & \equiv & \int^{a_0}_{a_{\rm{src}}} \frac{{\rm{d}} a}{a^2 H} ~, \\
		B & \equiv & \frac{Z - Z_{\rm{src}}}{\sqrt{1 + \upsilon^2}} + \upsilon
		\int^{a_0}_{a_{\rm{src}}} \frac{{\rm{d}} a}{a^2 H \sqrt{a^2 + \upsilon^2}}
		~.
	\end{eqnarray} 
\end{subequations}
The $Z_1$ and $Z_2$ in the integration in Eq.~(\ref{40}) can be determined by
solving the equation of the light ray along $Z$-axis, 
\begin{equation}
  \frac{{\rm{d}} Z_{\pm}}{{\rm{d}} T} = \pm \sqrt{\frac{1 + \upsilon^2}{a^2 +
  \upsilon^2}} ~. 
\end{equation}
Considering initial condition of the above equation $Z_{\pm} (T_{\rm{src}}) = Z_{\rm{src}}$, we can obtain $Z_1 \equiv Z_- (0)$ and $Z_2 \equiv Z_+ (0)$. 

By substituting Eq.~(\ref{40}) into Eq.~(\ref{29}), we can obtain the $d_L$ expressed in terms of  $T_{\rm{src}}$ and $Z_{\rm{src}}$. The value of luminosity distance depends on both the distance ($T_{\rm src}$) and the location ($Z_{\rm src}$) of the celestial objects (sources). On the other side, in order to obtain luminosity distance-redshift relation, we should also consider the redshift of the sources. Based on Eqs.~(\ref{38}) and (\ref{39}), in the case that a light ray emitted at event $X^{\mu}_{\rm{src}}$ pass through the event $X^{\mu}= 0$ for our observers, the $\gamma$ can be expressed in terms of $T_{\rm{src}}$ and $Z_{\rm{src}}$, i.e., 
\begin{eqnarray}
  \gamma & = & \left( - \frac{Z_{\rm{src}}}{\sqrt{1 +
  \upsilon^2}} + \upsilon \int^1_{a_{\rm{src}}} \frac{{\rm{d}} a}{a^2 H
  \sqrt{a^2 + \upsilon^2}} \right) \left( \int^1_{a_{\rm{src}}} \frac{{\rm{d}}
  a}{a^2 H} \right)^{- 1} ~. \label{41}
\end{eqnarray}
In this sense, we can ensure that the redshift in Eq.~(\ref{22}) can also be expressed in terms of $T_{\rm{src}}$ and $Z_{\rm{src}}$. Therefore, from Eqs.~(\ref{22}), (\ref{29}) and (\ref{40})--(\ref{41}), the parametric equations of luminosity distance-redshift relation can be rearranged formally as
\begin{subequations}
	\begin{eqnarray}
		d_L  &=  &(1 + z) \sqrt{\frac{\mathcal{S}(T_{\rm src}, Z_{\rm src})}{4 \pi}} ~,\\
		z 	& = & \frac{1}{a_{\rm{src}}^2} \frac{\sqrt{a_{\rm{src}}^2 + \upsilon^2}
			- \upsilon \gamma(T_{\rm src}, Z_{\rm src})}{\sqrt{1 + \upsilon^2} - \upsilon \gamma(T_{\rm src}, Z_{\rm src})} -1~.
	\end{eqnarray}\label{42}
\end{subequations}
Based on Eqs.~(\ref{23}) and (\ref{41}), we can also rewrite the expression of the observed location $\Theta$ from the $Z$-axis for the sources,
\begin{equation}
	\Theta = \arccos \left(- \frac{\gamma(T_{\rm src}, Z_{\rm src}) \sqrt{1 + \upsilon^2} - \upsilon}{\sqrt{1 +	\upsilon^2} - \upsilon \gamma(T_{\rm src}, Z_{\rm src})} \right)~.
\end{equation}

So far, above derivations are independent on cosmological model. In the following, we will present the luminosity distance-redshift relations in two specific cases, namely, de Sitter model and $\Lambda$CDM model.

\subsubsection{de Sitter Universe}

In de Sitter Universe, the scale factor is given by $a =  e^{H_0 (t -t_0)}$. It can describe  dark energy dominated Universe in low redshift regime. Since the form of scale factor is given, we can express the $a$ in terms of coordinate system $(T, X, Y, Z)$ by making use of Eq.~(\ref{4}),
\begin{equation}
  a = \frac{1}{2} \left( \left( 1 + \sqrt{1 + \upsilon^2} \right) e^{H_0
  \left( T + \frac{\upsilon Z}{\sqrt{1 + \upsilon^2}} \right)} + \left( 1 -
  \sqrt{1 + \upsilon^2} \right) e^{- H_0 \left( T + \frac{\upsilon Z}{\sqrt{1
  + \upsilon^2}} \right)} \right) ~. \label{41-1}
\end{equation}
Based on Eqs.~(\ref{39}) and (\ref{41-1}), we obtain equation of isochronous surface for the light rays from a source at event $(T_{\rm src}, X_{\rm src}, Y_{\rm src}, Z_{\rm src})$,
\begin{eqnarray}
  \left( \frac{H_0 (Z - Z_{\rm{src}})}{\sqrt{1 + \upsilon^2}} +
  \frac{1}{\upsilon} \left( \sqrt{1 + \left( \frac{\upsilon}{a} \right)^2} -
  \sqrt{1 + \left( \frac{\upsilon}{a_0} \right)^2} \right) \right)^2 \nonumber \\
  {}  +  H_0^2
  (X - X_{\rm{src}})^2 + H_0^2 (Y - Y_{\rm{src}})^2 & = & \left(
  \frac{1}{a_{\rm{src}}} - \frac{1}{a_0} \right)^2 ~.
\end{eqnarray}
It leads to the area of the isochronous surface $\mathcal{S}$ in Eq.~(\ref{40}) with
\begin{subequations}
\begin{eqnarray}
	A & = & \frac{1}{H_0} \left( \frac{1}{a_{\rm{src}}} - \frac{1}{a_0}
	\right) ~, \\
	B & = & \frac{Z - Z_{\rm{src}}}{\sqrt{1 + \upsilon^2}} + \frac{1}{\upsilon
		H_0} \left( \sqrt{1 + \left( \frac{\upsilon}{a_{\rm src}} \right)^2} - \sqrt{1 +
		\left( \frac{\upsilon}{a_0} \right)^2} \right) ~.
\end{eqnarray}\label{44}
\end{subequations}
Substituting Eqs.~(\ref{44}) into Eq.~(\ref{29}), we can obtain the luminosity distance-redshift relation in form of Eq.~(\ref{42}).

\begin{figure}
	{\includegraphics[width=.6\linewidth]{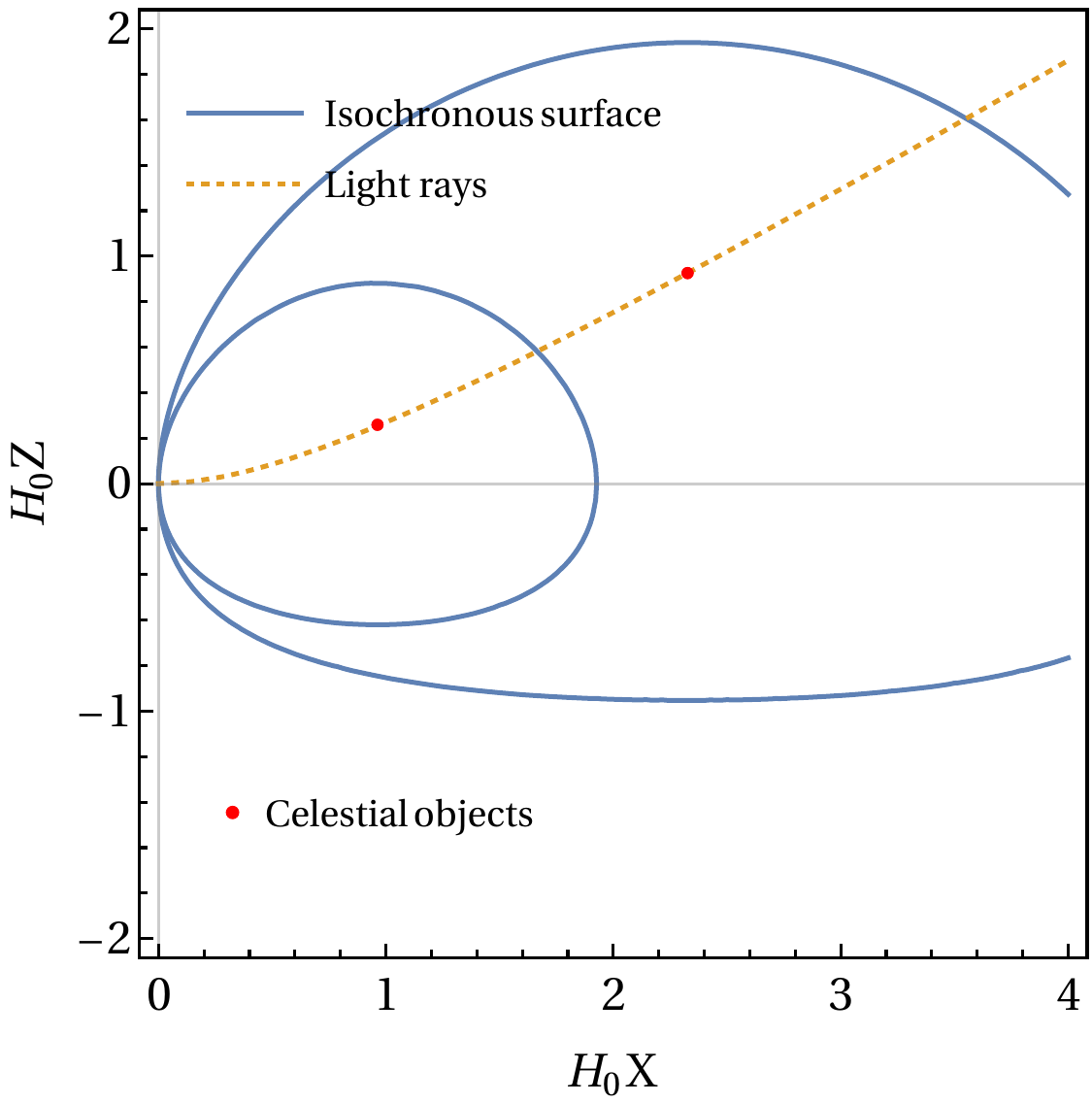}
	\caption{Schematic diagram for two sources (celestial objects) at observed location $\Theta=90^\circ$. The observers are located at the origin of the coordinate system $(X, Z)$. The dashed line represent the light rays from the sources. The solid curves represent the isochronous surface projected in the $X$-$Z$ plane. For illustration, we select a quite large value parameter 	$\upsilon$.}\label{F2}}
\end{figure}

\begin{figure}
	{\includegraphics[width=1\linewidth]{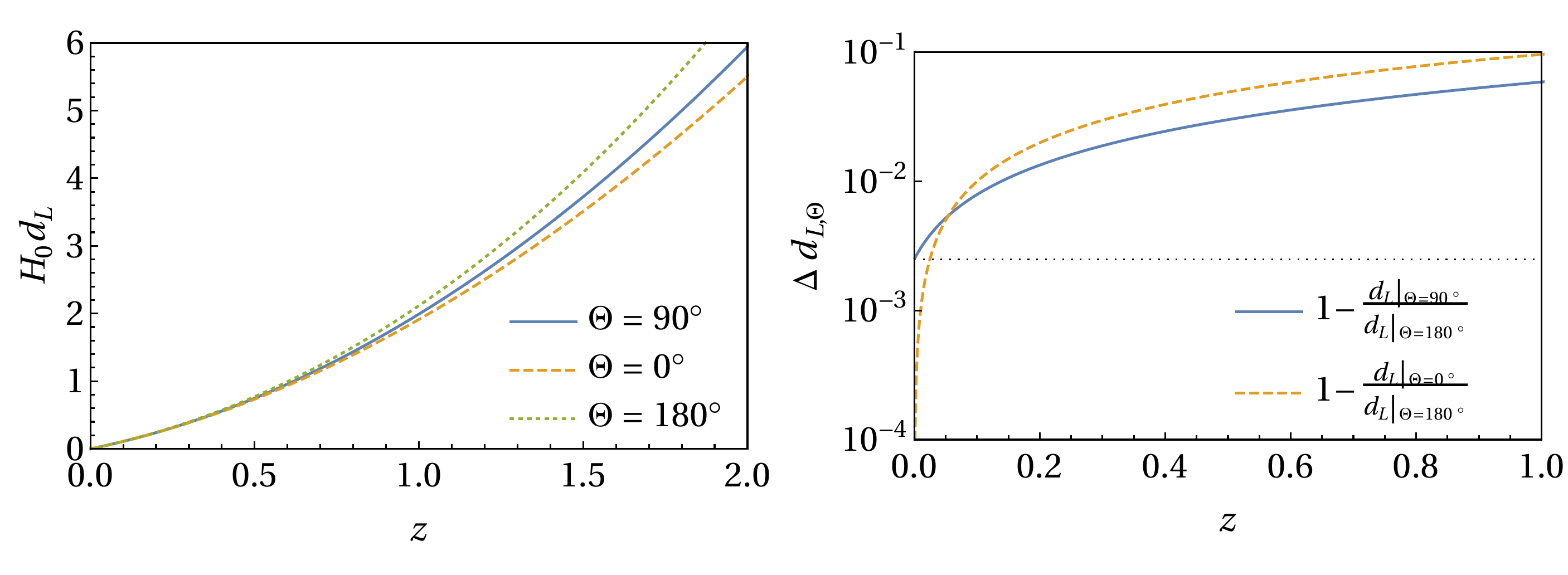}
	\caption{Left panel: Luminosity distance-redshift relation for selected observed location $\Theta$. Right panel: The value of $\Delta d_{L, \Theta} $ as function of redshift. In both panels, the parameter $\upsilon$ in the anisotropic FLRW space-time is set as 0.05. The Hubble parameter $H$ is a constant in de Sitter Universe.}\label{F3}}
\end{figure}

In Figure~\ref{F2}, we show schematic diagram for sources at observed location $\Theta=90^\circ$. Due to the bending of light rays, these sources, in fact, exist in the Universe with the co-moving coordinates $Z_{\rm src}>0$. Besides, one might find that the isochronous surface is no longer a sphere. 
In right panel of Figure~\ref{F3}, we present luminosity distance-redshift relations in three different locations $\Theta$.  In the regime of $z \simeq 1$, the
Universe tends to have a larger expanding rate in the direction along the
geodesic motion of reference frame, namely, $\Theta = 0^{\circ}$, while the Universe has the smallest
expansion rate in the direction, $\Theta = 180^{\circ}$.
In order to study difference of luminosity distance in different locations $\Theta$ for given redshift $z$, we introduce a quantity $\Delta d_{L, \Theta}$ in the form of
\begin{equation}
  \Delta d_{L, \Theta} \equiv 1- \frac{d_{L}\big|_{ \Theta}}{d_{L}\big|_{ \Theta_0=180^\circ}}  ~.
\end{equation}
In right panel of Figure~{\ref{F3}}, we plot the curves of $\Delta d_{L,\Theta}$ as function of redshift $z$. It shows that  the expansion rate at the location $\Theta=180^\circ$ is always less than that at the locations $\Theta=0^\circ$ and $\Theta=90^\circ$.  In low redshift limit, we have
\begin{eqnarray}
 \Delta d_{L, 90^\circ} \Big|_{z = 0} &=& 1-\frac{\frac{z}{H_0\sqrt{1 + \upsilon^2}}}{\frac{z\sqrt{1 + \upsilon^2}}{H_0}}\Bigg|_{z=0} = \frac{\upsilon^2}{1 + \upsilon^2}~, \label{50}\\
  \Delta d_{L, 0^\circ} \Big|_{z = 0}& =& 1-\frac{\frac{z\sqrt{1 + \upsilon^2}}{H_0}}{\frac{z\sqrt{1 + \upsilon^2}}{H_0}}\Bigg|_{z=0} = 0~, \label{51}
\end{eqnarray}
which has been studied in Ref.~\cite{chang_reference_2020}. For the result in Eq.~(\ref{50}), we have plotted the dotted line in right panel of Figure~{\ref{F3}}. From Eq.~(\ref{51}), it indicates that the expansion rates in  positive and negative direction of $Z$-axis are shown to be the same in the case of $z \rightarrow0$.  More interestingly, we notice that there is a transition point at $z\simeq \upsilon$ for the luminosity distance-redshift relation. Namely, the expansion rate at location $\Theta=0^\circ$ turns to be larger than  the expansion rate at location $\Theta=90^\circ$ in the regime of $z \gtrsim \upsilon$. 

\begin{figure}
	{\includegraphics[width=1\linewidth]{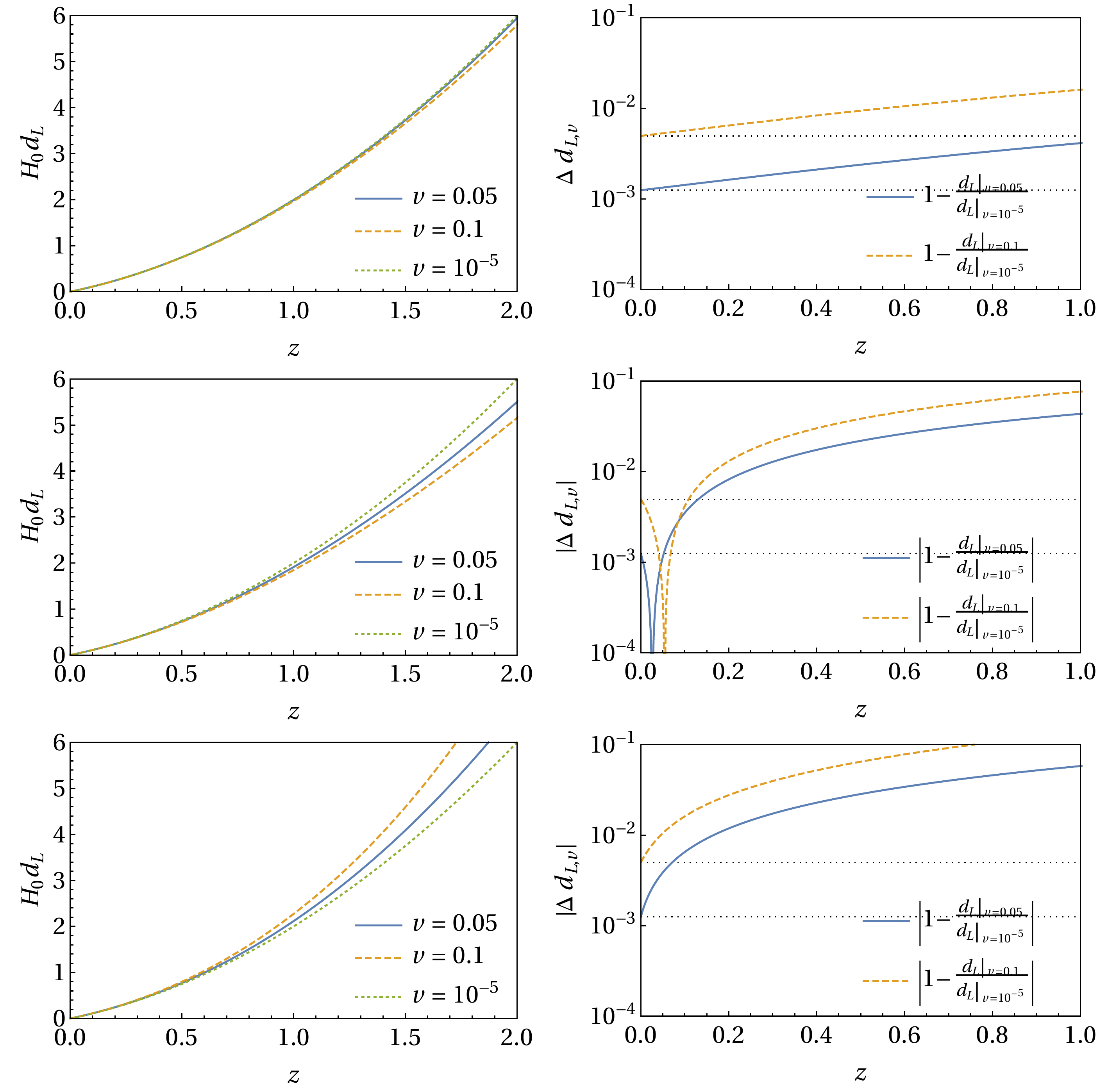}
	\caption{Left panel: Luminosity distance-redshift relation for selected parameter $\upsilon$. Right panel: The value of $\Delta d_{L, \upsilon} $ as function of redshift. In the top, middle and bottom panels, we set $\Theta=90^\circ$, $0^\circ$ and $180^\circ$, respectively. The Hubble parameter $H$ is a constant in de Sitter Universe.}\label{F4}}
\end{figure}

In left panel of Figure~\ref{F4}, we plot the luminosity
distance-redshift relation for selected parameters $\upsilon$. The expansion rate at the location $\Theta=90^\circ$ tends to be insensitive to the parameter $\upsilon$.
In right panel of Figure~\ref{F4}, we present difference of the luminosity distance as function of redshift for selected parameters $\upsilon$, which is defined as
\begin{equation}
	\Delta d_{L, v} \equiv 1- \frac{d_{L}\big|_{ v}}{d_{L}\big|_{ v=10^{-5}}}  ~.
\end{equation}
In low redshift limit, the dotted lines in the plots can be obtained by making use formulae in Ref. \cite{chang_reference_2020}. In the mid-right panel of Figure~\ref{F4}, we also notice the  transition points at $z\simeq \upsilon$ for the luminosity distance-redshift relation.

\begin{figure}
	{\includegraphics[width=.6\linewidth]{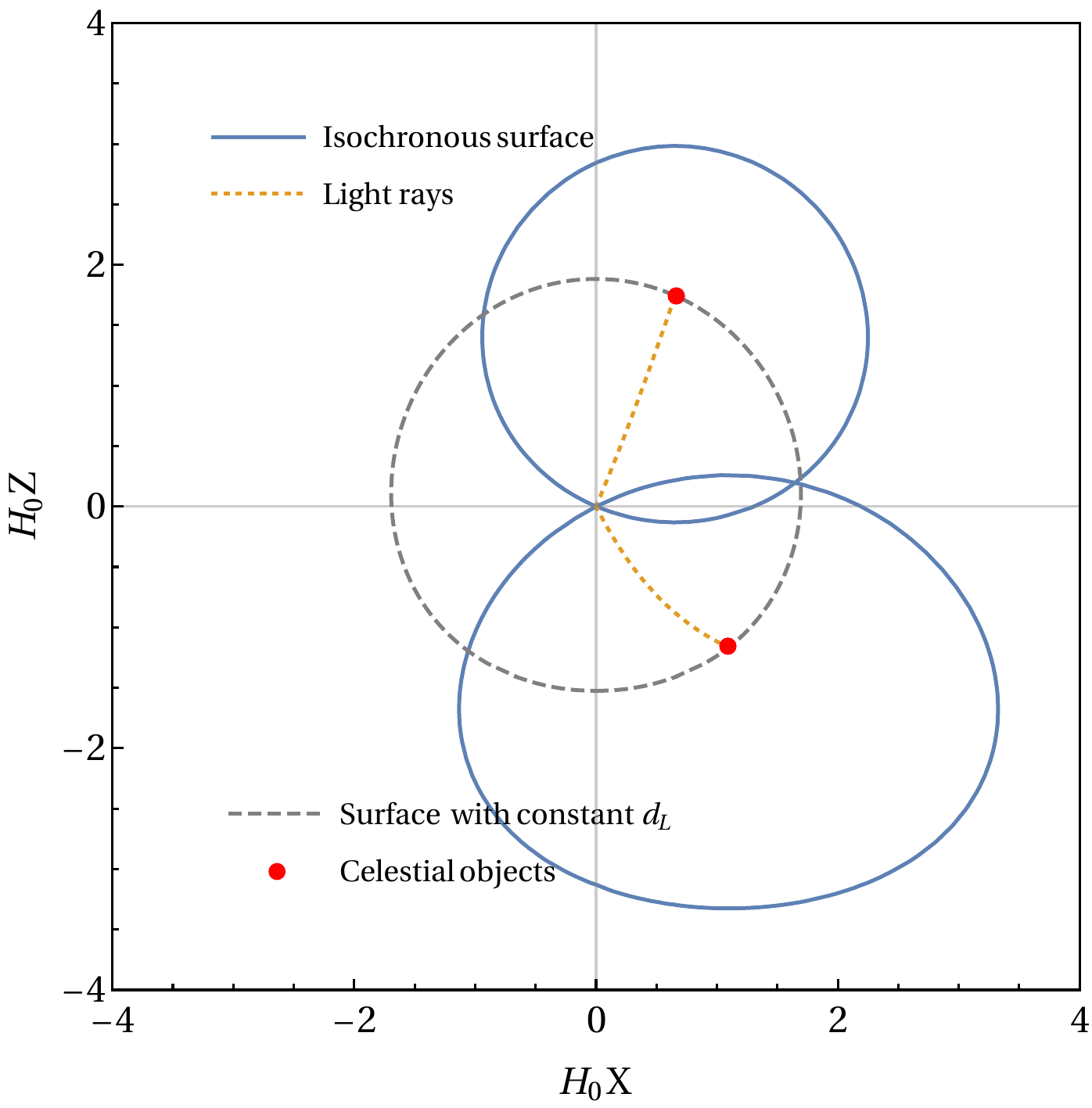}
	\caption{ Schematic diagram of the spatial surface (dashed curve) with constant $d_L$. We plot two sources (celestial objects) for example. The observers are located at the origin of the coordinate system $(X, Z)$. The dashed line represent the light rays from the sources. The solid curves represent the isochronous surface projected in the $X$-$Z$ plane. For illustration, we select a quite large value of parameter 	$\upsilon$.}\label{F5}}
\end{figure}

\begin{figure}
	\includegraphics[width=1\linewidth]{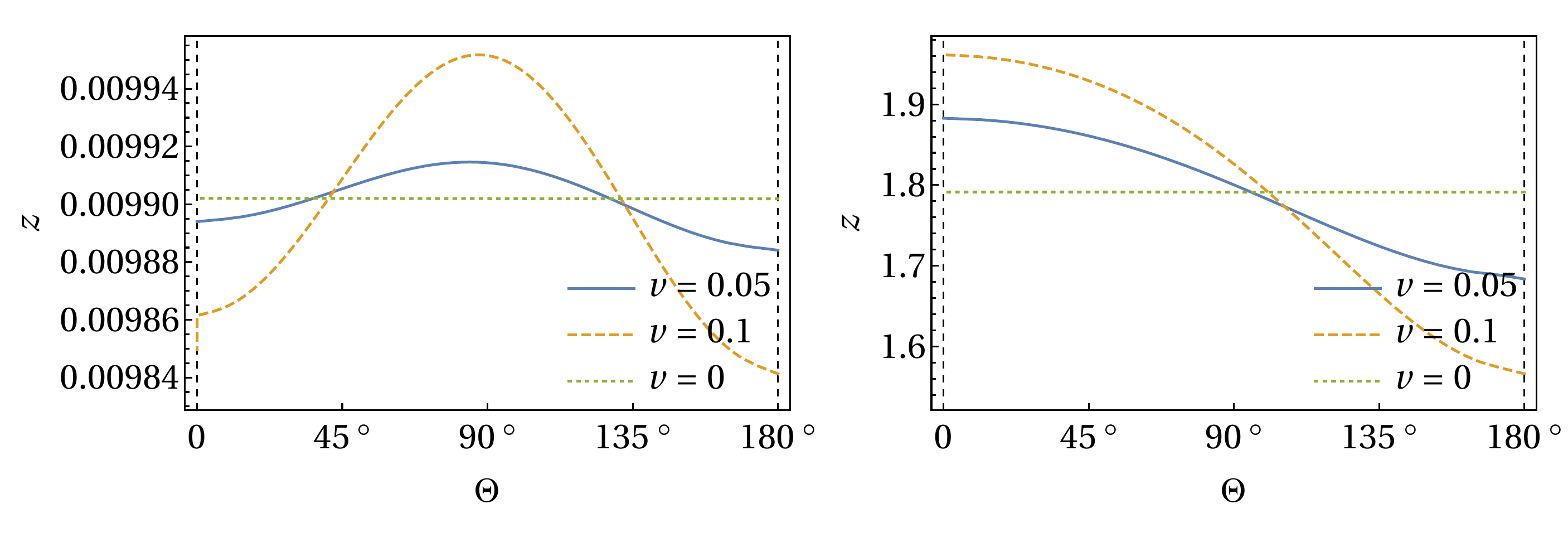}
	\caption{Redshift as function of observed location of sources $\Theta$ for selected parameters $\upsilon$, and luminosity distance, $d_L =0.01H_0^{-1}$ (left panel) and $d_L =5H_0^{-1}$ (right panel). The Hubble parameter $H$ is a constant in de Sitter Universe.}\label{F6}
\end{figure}

Besides, we  could also consider the redshift in a spatial surface with constant $d_L$. The schematic diagram is shown in Figure~\ref{F5}. The shape of isochronous surfaces could be different for the sources at different location $\Theta$.
In Figure~\ref{F6}, we present redshift as function of observed location $\Theta$ of sources  for given luminosity distance. In the case of $\upsilon \neq 0$, the observed redshift is shown to  vary with location of the source. In the regime of $z > \upsilon$, the curves of $z(\Theta)$ tend to be monotonic, while in the regime of $z < \upsilon$, the redshift would get the largest value at the location $\Theta\simeq90^\circ$.

Therefore, based on these results shown in Figure~\ref{F2}--\ref{F6}, we could conclude that in the regime of $z > \upsilon$, the luminosity distance-redshift relation tends to have a dipole structure, while in the regime of $z \ll \upsilon$, there is not such dipole. 

\subsubsection{$\Lambda$CDM Universe}

In $\Lambda$CDM model, the Hubble parameter can be given by the Friedman
equation,
\begin{eqnarray}
  H & = & H_0 \sqrt{\Omega_m (1 + z)^3 + \Omega_{\Lambda}} ~, \label{51-1}
\end{eqnarray}
where $\Omega_{\Lambda} + \Omega_m = 1$. 
In this case, we could not obtain an analytic expression of scale factor $a$. Thus, we turn to study the luminosity distance-redshift relation with numerical method.

\begin{figure}
	{\includegraphics[width=1\linewidth]{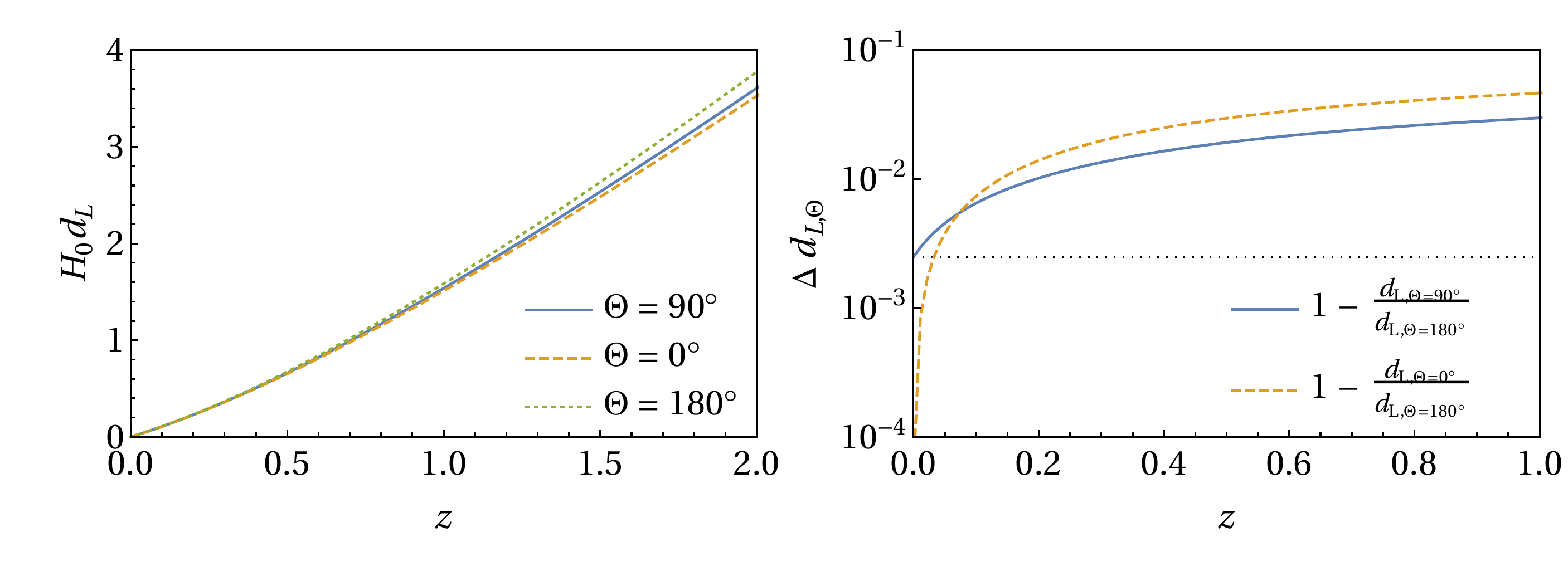}
	\caption{Left panel: Luminosity distance-redshift relation for selected observed location $\Theta$. Right panel: The value of $\Delta d_{L, \Theta} $ as function of redshift. In both panels, the parameter $\upsilon$ in the anisotropic FLRW space-time is set as 0.05. The Hubble parameter $H$ is given in Eq.~(\ref{51-1}).}\label{F7}}
\end{figure}

\begin{figure}
	{\includegraphics[width=1\linewidth]{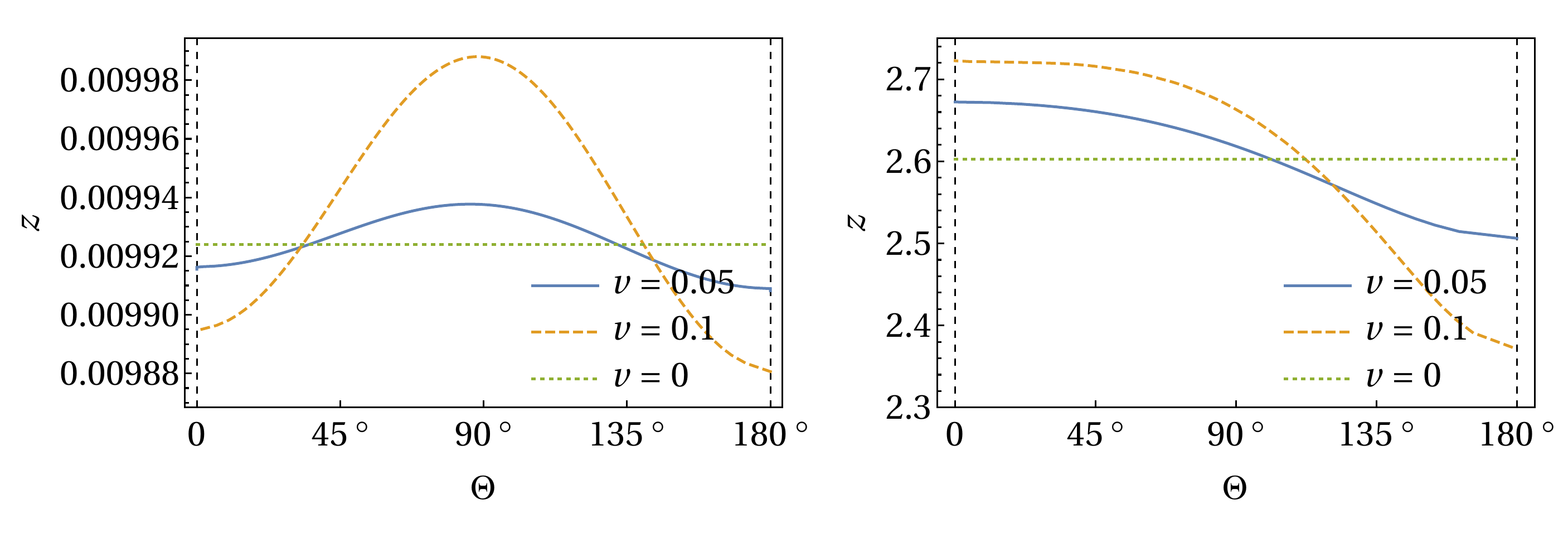}
	\caption{Redshift as function of observed location of sources $\Theta$ for selected parameters $\upsilon$, and luminosity distance $d_L =0.01H_0^{-1}$ (left panel) and $d_L =5H_0^{-1}$ (right panel). The Hubble parameter $H$ is given in Eq.~(\ref{51-1}).}\label{F8}}
\end{figure}

For given observed location of the sources, we present luminosity distance-redshift relation in Figure~\ref{F7}. It is similar to the results in de Sitter Universe. Namely, in the case of $\Theta = 0^\circ$, the expanding rate tends to be the largest, while in the case of $\Theta = 180^{\circ}$, the expanding rate tends to be the smallest.
For given  luminosity distances, we plot the curves of redshift as function of location $\Theta$  in Figure~\ref{F8}.
Similarly, in the regime of $z > \upsilon$, the curve of $z(\Theta)$ tends to be monotonic, while in the regime of $z < \upsilon$, the curve of $z(\Theta)$ is shown to be unimodal. The spatial dipole structure in high redshift regime also exists in $\Lambda$CDM model.

\begin{figure}
	{\includegraphics[width=.7\linewidth]{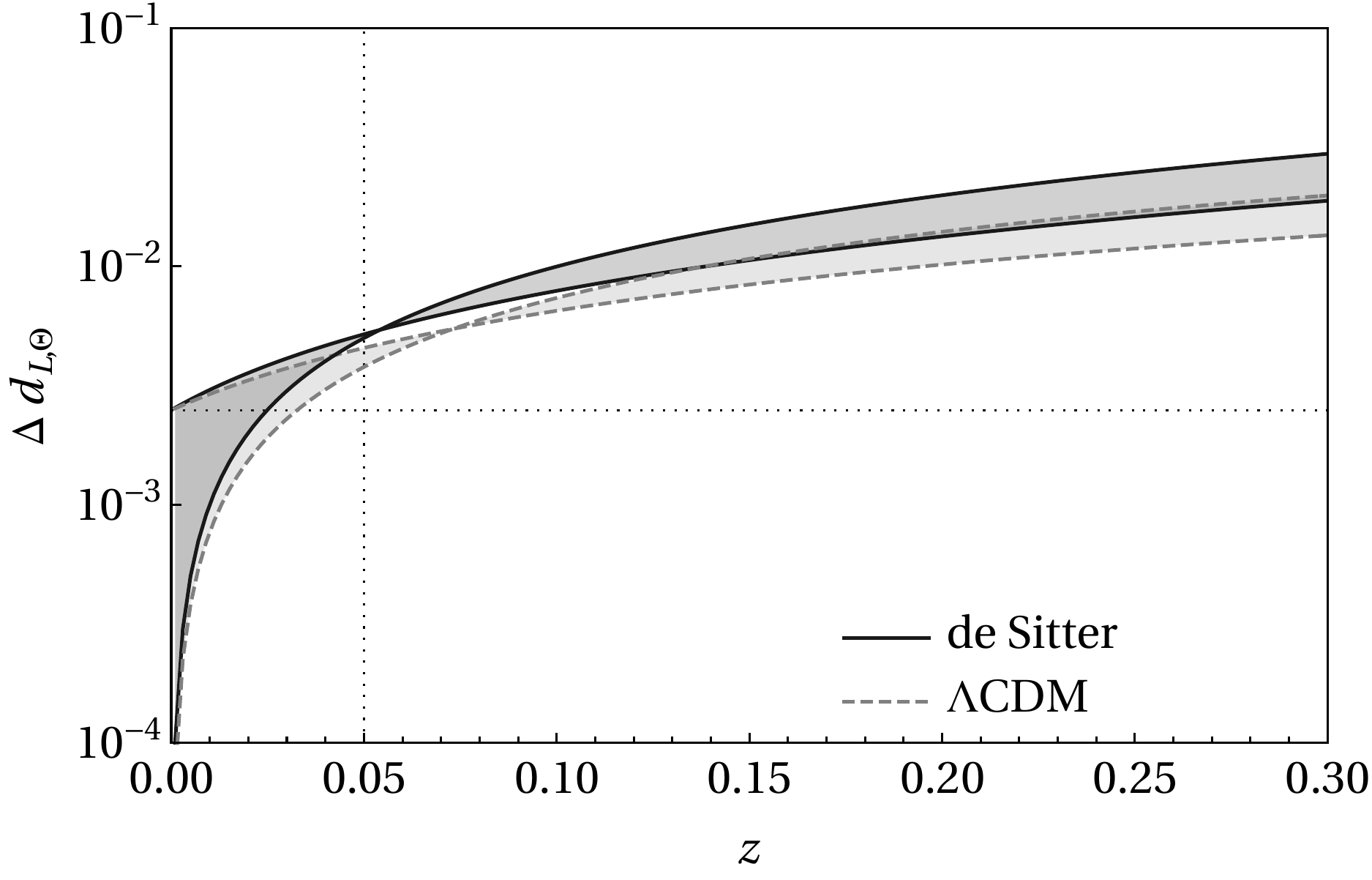}
	\caption{A comparison of the curves of $\Delta d_{L,\Theta}(z)$ between the results $\Lambda$CDM model in right panel of Figure~\ref{F7} and the results in the de Sitter model in right panel of Figure~\ref{F3}.}\label{F9}}
\end{figure}

Figure~\ref{F9} shows a comparison of the curves of $\Delta d_{L,\Theta}(z)$ between the results $\Lambda$CDM model and the results in the de Sitter model. In the low redshift regime, the luminosity distance-redshift relation is  less sensitive to cosmological models. In the regime of $z\simeq\upsilon$, the transition points are shown to be not precisely at $z=\upsilon$. It depends on cosmological models in general.

\section{Conclusions and Discussions} \label{V}

In this paper, we explored the possibility that the spatial variation of the fine-structure constant could be compatible with Einstein's general relativity. Namely, the value of $\Delta \alpha / \alpha$ in the Universe is originated in different values of the speed of light in separate local frames. In this sense, we found that the spatial variation of the fine-structure constant could be deduced, if defining the non-local measurement of speed of light $c_z$ in Eq.~(\ref{9}) and use it in the anisotropic FLRW space-time. Besides, we also studied the luminosity-distance relation in the anisotropic FLRW space-time. It is found that there is a dipole structure in high redshift regime, while there is not such dipole in low redshift regime. 

We showed that the phenomenological formulae in Eqs.~(\ref{7}) and (\ref{9}) is acceptable based on the application in Milne Universe. In physical view-point, Eq.~(\ref{9}) indicates how we define the non-local measurement of speed of light $c_z$. Thus, it is expected that the phenomenological study on $c_z$ might improve our understanding of the non-local measurement in general relativity. 


The variation of the fine-structure constant can also be interpreted as a varying charge or Planck constant in the Universe \cite{magueijo_is_2002}. We thus expect that a more rigours study should be based on quantum field theory or quantum mechanics in curved space-time, in which the $c_z$ (or $e_z$, \emph{etc}.) might be obtained based on the first principle. 

The anisotropic coordinate of FLRW space-time was originally proposed aiming for relieving the Hubble tension \cite{chang_reference_2020}. It concluded that the parameter $\upsilon$ is inferred to be $\upsilon\lesssim0.25$ based on redshift survey at the regime of $z<0.03$  \cite{qin_bulk_2019}. However, in this work, we obtained $\upsilon\simeq10^{-5}$ at the regime of  redshift $z> 0.2$ based on the quasar spectra \cite{king_spatial_2012}. The different values of $\upsilon$ suggests that the $\upsilon$ might not be universal constant and should vary with the redshift $z$. 

\acknowledgments
The authors wish to thank Prof.~Sai Wang for discussions.  This work has been funded by the National Nature Science Foundation of China  under grant No. 11675182, 11690022 and 12075249.

\bibliography{cite}
\end{document}